\documentclass[journal]{IEEEtran}
\usepackage{graphicx}
\usepackage[normalem]{ulem}
\usepackage{amssymb}
\usepackage{amsmath}
\usepackage{amstext}
\usepackage{amsfonts}
\usepackage[normalem]{ulem}
\usepackage[english]{babel}
\usepackage[table,xcdraw]{xcolor}
\usepackage{epstopdf}
\usepackage{amsmath}
\usepackage{caption}
\usepackage{subcaption}
\usepackage{amssymb}
\usepackage{url}
\usepackage{hyperref}

\usepackage{mathtools}
\usepackage{arydshln}
\usepackage{pstricks}

\renewcommand\thetable{\thesection.\arabic{table}}

\begin{document}
\renewcommand*\abstractname{Abstract}
\renewcommand{\figurename}{Figure}
\renewcommand{\tablename}{Table}
\renewcommand{\refname}{References}

\title{A Vis-Spectrophotometer with a new topology of nanophotonic diffraction grating}
\author{{\textbf{Victor Vermehren Valenzuela}, vvalenzuela@uea.edu.br \\
State University of Amazon, UEA, Av. Darcy Vargas 1200, Manaus, AM, Brazil\\
\vspace{0.3 cm}
\textbf{H. M. de Oliveira}, hmo@de.ufpe.br}
\\ Federal University of Pernambuco, UFPE, Av. Prof Moraes Rêgo 1235, Recife, Brazil}

\maketitle

\begin{abstract}
\textbf{
This paper presents a new topology for photonic crystals to replace the monochromator, introducing them with a new alignment (the so-called extended Trinitron) to guide the chromatic pattern producing a spatial distribution similar to that of conventional diffraction gratings. The spectrophotometer uses a LED white light as light source instead of chambers with halogen lamps and mirrors. The system was designed using actual data from the manufacturer of photonic crystals and the total system response is compared against that one of a conventional spectrophotometer with a LED white light or tungsten bulb as a light source. It is shown that the spectral responses are similar in photonic panels containing more than twenty elements.
} 
\end{abstract}

\providecommand{\keywords}[1]{\textbf{\textit{Index terms---}}}
\section{Introduction}
\vspace{0.3 cm}
In recent decades, the miniaturization of spectrometers has experienced a major breakthrough with the advent of the SMD technology (Surface Mounted Device), which made possible the compacting of printed circuit boards. Nevertheless, little has changed in conventional optical components: the diffraction gratings, in which the relationship between diffraction angle and wavelength directly affects the dimensions of the monochromator and the resolution of the spectrophotometer reading; mirrors, whose durability is the humidity conditioned and clean; and light sources, with its intrinsic high degree of heat emission, and also high operating currents. Here, we present a new topology that allows minimizing the spectrophotometer by solving these problems and by eliminating the main drawbacks. The conventional spectrophotometer performs spectral measurements by reading the projection of light chromatic diffracted by diffraction gratings. The idea here was to reproduce that same formatting geometric projection that differentiated the wavelengths of light input: a base formed by a photonic crystal (PhC) pattern - which is here termed as Trinitron pattern. A separate light source and away from the grid for reasons of heat and angle of diffraction can be characterized by a LED of high brightness white led the panel of photonic crystals.\\

The state of the art of the miniaturization of optical spectrometers can be seen in \cite{Wolffenbuttel}, \cite{Avrutsky_et_al}, as well as details on their portability \cite{UVOIRSpectroscop}, \cite{Berg_et_al}. Other articles propose techniques for miniaturization, but with a focus on certain sections of the spectrophotometer, e.g. the replacement of the CCD detector \cite{Adams_et_al}, the introduction of diffraction grating with transmission in the waveguide \cite{Grabarnik_et_al}, \cite{Sander_et_al}. Topologies with geometry rather similar to that addressed in this article can be found in \cite{Momeni_et_al}, where is introduced a photon prism \cite{Webster} using an array of LEDs, and \cite{Pervez_et_al}, which proposes the projection chromatic of light. In the latter, the proposed geometry precludes the practical application due to the selectivity of the absorbance measured by the spectrophotometer. The equipment presented here operates in the visible band and revealed to be low cost. It is extremely simple, and property-based in line photonic crystals that selectively couple the guided wavelengths of light so as to simulate the diffraction grating.

\section{Principle of Operation of the Spectrophotometer}

Photonic crystals are periodic dielectric media with low losses and minimal absorption of light, in which the light refractions and reflections from these interfaces present the same phenomena as the photon in the atomic level \cite{Joannopoulos_et_al}. Thus, the optical control and manipulation can be determined by the choice of the crystal structure, the shape and size of gaps, the thickness of the dielectric layer, the variation of $\epsilon$ (permittivity) and the lattice constant $a$. The photonic crystal structure used here is similar to the structure of light extraction from photonic light emitting diodes - PhC LED \cite{Ishihara_et_al}. Other schemes, involving the coupling of guided light using photonic crystals can be found in \cite{Philips}, \cite{Osram}.\\

\begin{figure}[!ht]
\centering
\includegraphics[width=9cm, height=4cm]{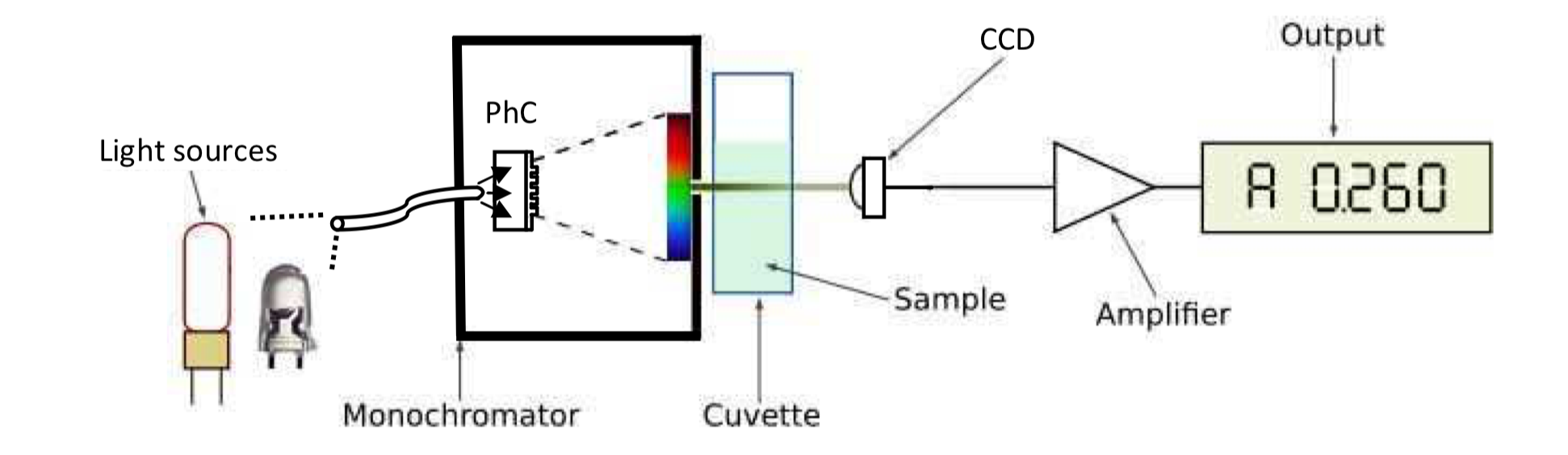}
\caption{Schematic operation of the spectrophotometer divided in the main blocks highlighting the introduction of photonic crystals panel}.
\label{fig:Schematic}
\end{figure}

Figure \ref{fig:Schematic} shows the operation of the spectrophotometer divided into its major blocks. The light from the selected source is transmitted through an optical fiber and then applied to the edges of a transparent substrate inside the monochromator. In the layer above the substrate are lithographed blocks of photonic crystals in the form of strips which can be geometric conformation of holes or rods with twenty different distances between dielectrics. Each of these strips of photonic crystals is designed in such a way that reflects twenty different spectrum bands. The choice that was based on a quantitative relationship between the total visible band (400 to 700 nm) and the FWHMs (Full Widths at Half Maximum) of the spectral response of each strip. The entire visible range is as a result covered without leaving gaps that could compromise the response of the spectrophotometer. The camera detector or CCD (Charge Coupled Device) is illuminated by the beam of light from photonic panel that selectively pass through the slot and you suffer losses from the sample into the cuvette containing the analyzed sample. The resulting electrical output of the CCD is compared with a reference and the difference is amplified and calibrated in such a way to characterize the photons absorbed by the sample. Note that the spectral response of the light sources will occur at different calibrations on the CCD to standardize the response.

Photonic crystals presented here are manufactured by lithography of poly methyl methacrylate (PMMA) on top of the glass substrate. The refractive index, nD, of PMMA at wavelengths of interest (400 to 700 nm) is between 1.499 and 1.512, slightly lower than that of glass ($n \approx 1.52$). The contrast medium is air. Photonic crystals are designed in a rectangular geometric pattern because of the extended Trinitron 20-strips of photonic crystals, with fill factor ($r / a$, where $r$ is the radius of the hole) of about 0.34. With such a parameterization, the periodic potential should not be intense enough to open up a whole band \cite{Ishihara_et_al}. The selectivity of the wavelengths emitted by the panel photonic engines can be implemented using high-precision step motors.

\section{Experiments and Simulations}

Blocks of PMMA (Microchem 495PMMA series) were lithographed on glass through the EBL (electron beam lithography) process, using glass slabs whose dimensions are 2.0 mm $\times$ 1.5 mm $\times$ 1.0 mm. The shape chosen for the structure was a hexagonal arrangement (fcc - face centered cube), which is easy to manufacture. The exposure was 320 $\mu C/cm^2$.\\
The determination of optimal values of the parameters of different lattice constants, filling factor, and thickness of PMMA were evaluated in the framework of light extracted from the crystal.
Thus, based on these figures, a panel of twenty rectangular crystalline structures with lattice constants from 245 to 425 nm and fill factor of 0.35 was chosen for a 400 nm PMMA thickness. Each spectrometer consists of blocks of 20 $\times$ 1 $\times$ 50 $\mu m$ 400 $\mu  m$ photonic crystal arrangement with a total of approximately 1.47 mm $\times$ 400 $\mu  m$.\\

The lighting system of photonic crystals was carried out using an optical fiber and a USB camera. The light was guided to the four edges of the substrate by a fiber optic cable of four branches (Lumitex Inc. and OptiLine). This assembly consists of a bundle of plastic fiber of 0.010", mounted in four segments of 0.80". Each segment was attached to one edge of the glass substrate.
RGB images of the system were taken using a USB camera (Matrix Vision mvBlueFox-120C) with an objective to increase and opening of 4X 0096 (Infinity Photo-Optical Infinistix). The spectral response was characterized using a programmable visible light source (FLP) from Horiba Tunable Light Source and Gemini 180 monochromator. The projection of photonic block was measured using a calibrated photodetector (Newport 918D).\\
Then, the performance of new photonic spectrophotometer was evaluated in comparison with a conventional optical spectrophotometer (HITACHI U-1100) with diffraction grating. First, replacing the entry by a 5-mm FLP ultra-white LED light \cite{Philips}, polarized in order to drive with a current of 5 mA. Afterward the LED was replaced by a tungsten lamp \cite{Osram}. The response was first characterized by using commercial spectrophotometer with its original tungsten lamp (Osram HALOSTAR Standard 64428), and then the lamp was replaced by a white LED light (Cool White LXK2-PW12-R00-Everlight), 5 mm, polarized the same way as in photonic spectrophotometer. Responses were analyzed through the SpectraSuite Spectrometer Operating Software supplied by Ocean Optics. Twenty regions of interest (ROI) of 8 $\times$ 32 pixels were defined for the system of 1 $\times$ 20 spectrophotometer. The luminance grayscale weighted for each ROI was calculated by averaging the values of red, green and blue in each region.

\section{Performance Evaluation}

\begin{figure}[!ht]
\centering
\includegraphics[width=9cm, height=4cm]{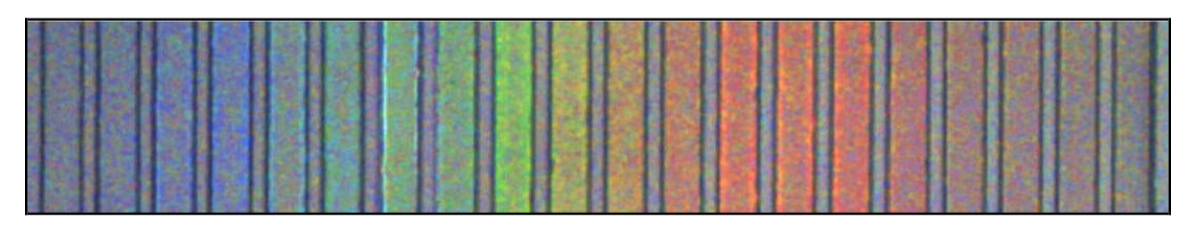}
\caption{The light emission of a white spectrum LED is coupled through the substrate to all segments of the photonic crystal panel.}
\label{fig:photonic_panel}
\end{figure}

\begin{figure}[!ht]
\centering
\includegraphics[width=9cm, height=4cm]{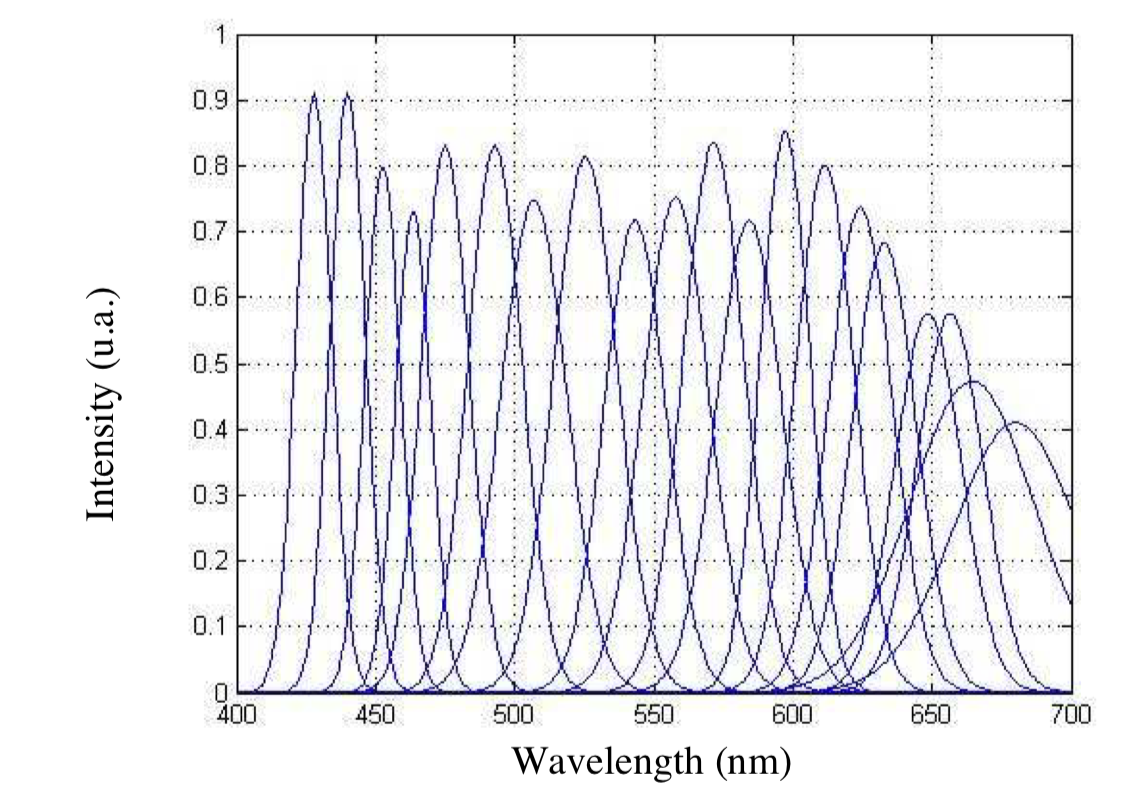}
\caption{Normalized spectral response of the 20X1 panel Trinitron standard photonic crystals, shown in Figure \ref{fig:photonic_panel}. Measurements represent the average intensity vs. wavelength. The peaks at each wavelength are evaluated by the lattice of each strip of PhC.}
\label{fig:spectral_response}
\end{figure}

Figure \ref{fig:photonic_panel} shows the optical microscopic image of the chromatic response of the block of twenty photonic crystals for an entry- spectrum white light (a tungsten halogen lamp). The regions illuminated with various colors in each of these strips illustrate the color variation of the photonic crystal responses, ranging from blue to shades of red, from left to right. The total set of images is used to calculate the luminance of the grayscale for each extended-Trinitron segment. Accordingly, the spectral response of each strip or basic functions is obtained, as shown in Figure \ref{fig:spectral_response}.

\begin{figure}[!ht]
\centering
\includegraphics[width=9cm, height=4cm]{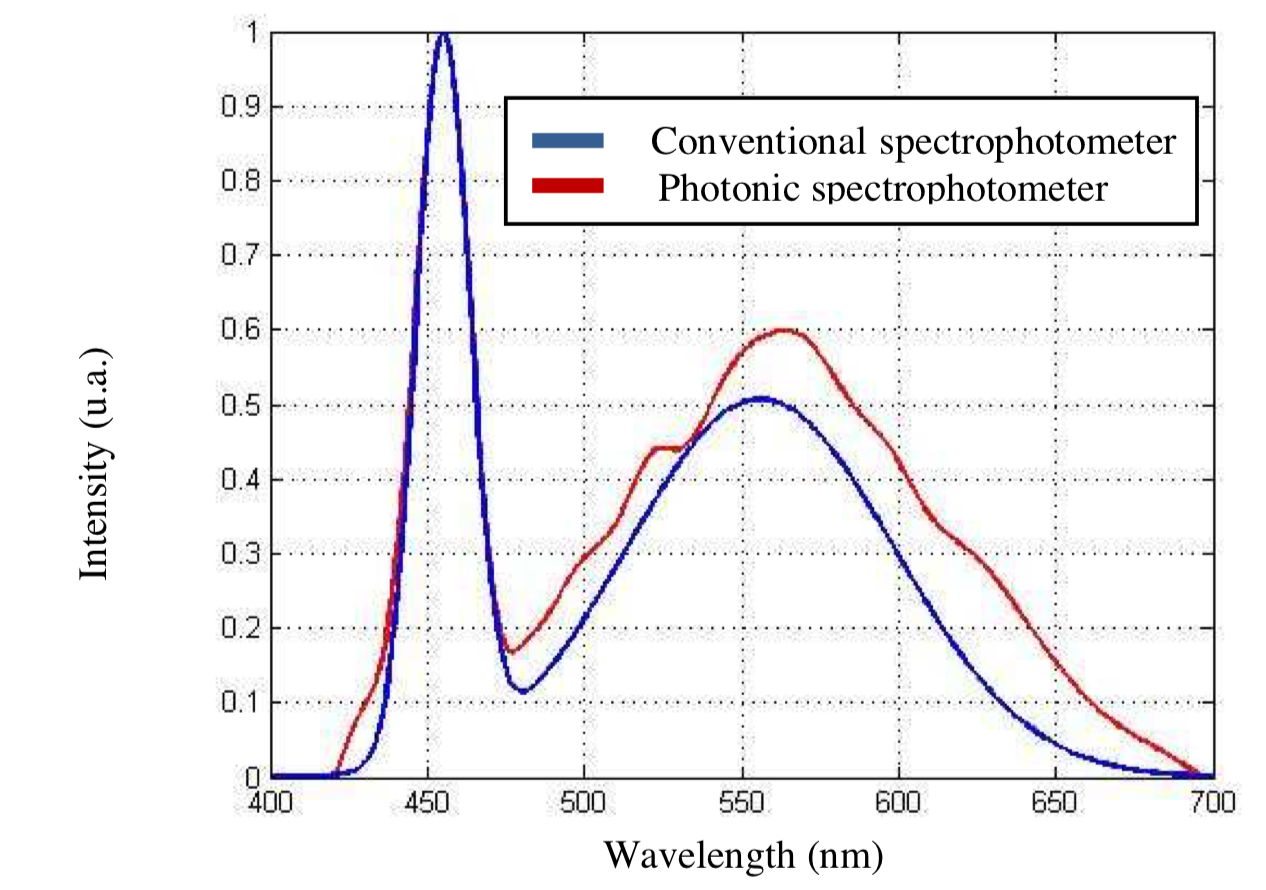}
\caption{Comparison between a photonic spectrometer and a conventional spectrometer (diffraction grating). Normalized curves exhibit the spectrum measured with white LED spectrum projecting on the diffraction grating of the conventional spectrophotometer and crystal panel of the photonic spectrophotometer.}
\label{fig:Comparison}
\end{figure}

The implementation presented here uses color images for information of the intensity pattern. The photonic spectrophotometer works by mapping the intensity of spatial patterns for the wavelength on the response of each element of the pattern Trinitron extended.\\
Figure \ref{fig:Comparison} shows the comparison of the performance between photonic spectrophotometer with twenty strips of Trinitron PhC standard and a conventional optical diffraction grating (HITACHI U-1100). The photon spectrum using the spectrometer was obtained through analysis of the grayscale intensities pattern of Figure  \ref{fig:photonic_panel}.

\begin{figure}[!ht]
\centering
\includegraphics[width=9cm, height=4cm]{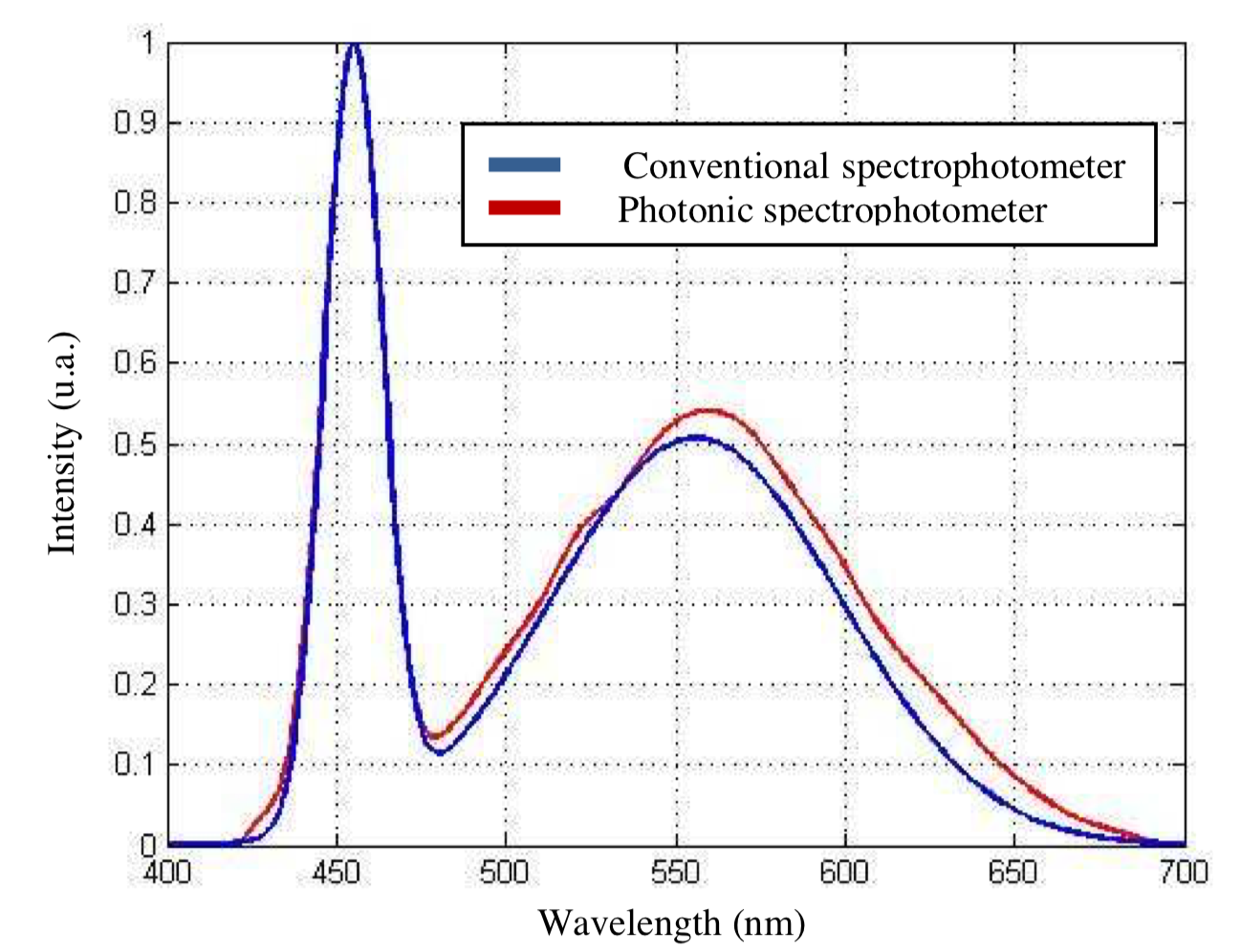}
\caption{Effect of number of base elements on the response of photonic spectrophotometer. Projection of the white spectrum LED on the substrate with 30 PhC strips.}
\label{fig:Effect}
\end{figure}

\begin{figure}[!ht]
\centering
\includegraphics[width=9cm, height=4cm]{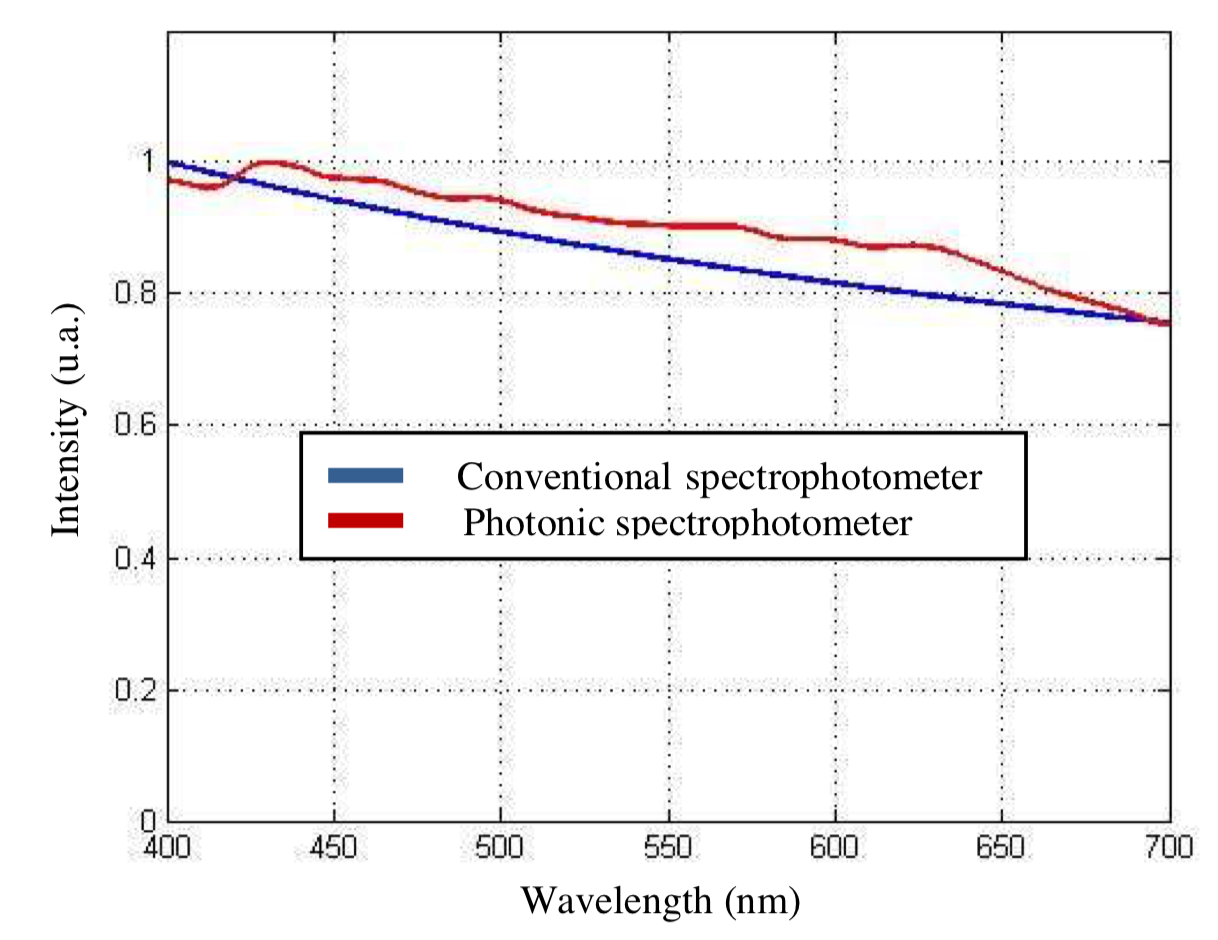}
\caption{Comparison between the photonic spectrophotometer and standard spectrophotometer (diffraction grating). Spectrum measured with the use of tungsten lamp in conventional spectrophotometer, and spectrum measured by projecting white light on the diffraction grating and the substrate of the photonic spectrophotometer.}
\label{fig:photonic_standard}
\end{figure}

\begin{figure}[!ht]
\centering
\includegraphics[width=9cm, height=4cm]{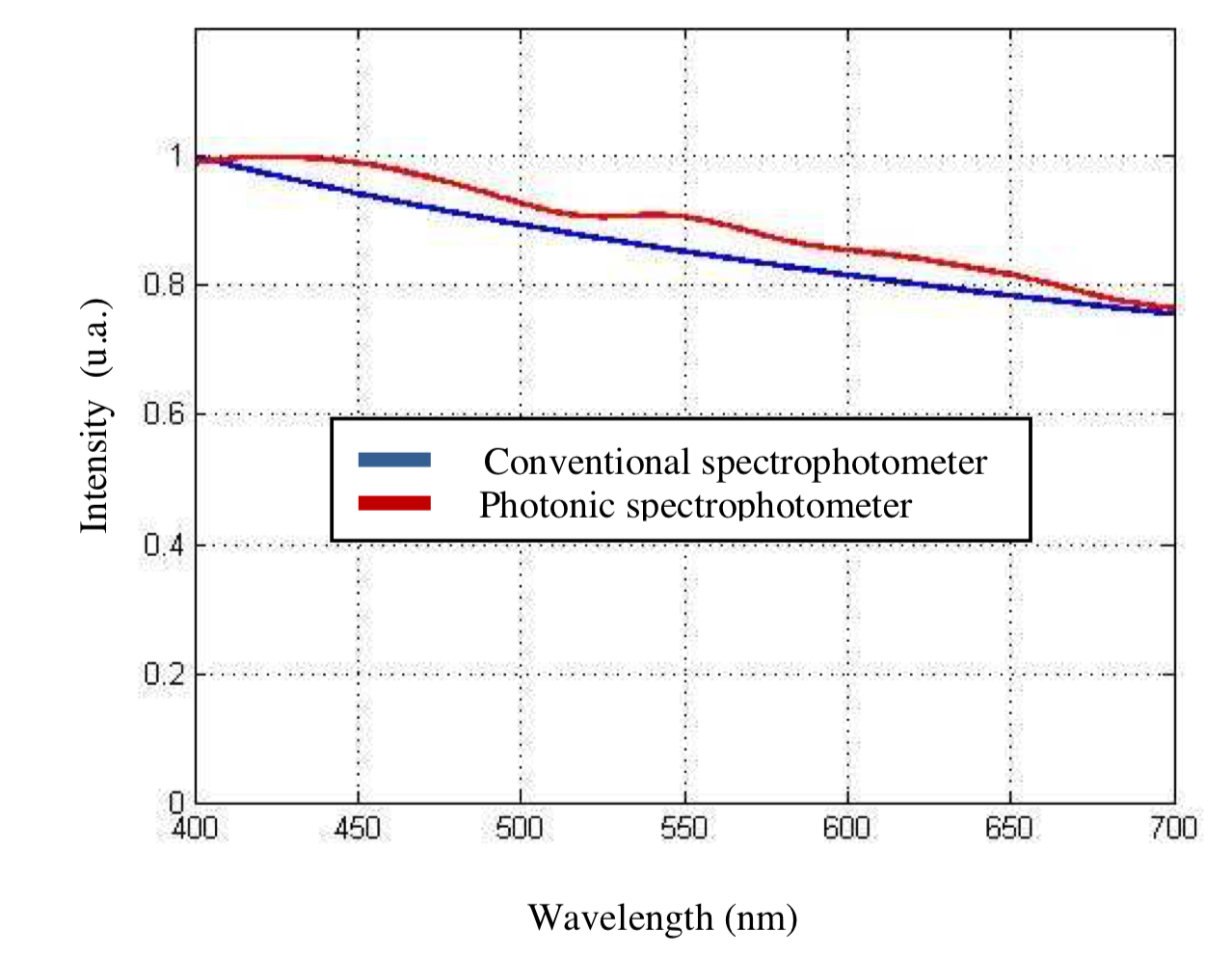}
\caption{Effect of number of base elements on the response of photonic spectrophotometer. Projection of the white spectrum LED on the substrate with 30 PhC strips.}
\label{fig:effect_base-elements}
\end{figure}

The blue spectrum, also shown in Figure \ref{fig:Comparison}, represents the limit of the system given the response pattern shown in Figure \ref{fig:spectral_response}. It is obtained by projecting the spectrum white LED (Cool-White LXK2-PW12-R00 Everlight) measured with the spectrophotometer on the commercial basis functions shown in Figure \ref{fig:spectral_response}.
The accuracy of the recovered spectrum can be improved by appending photonic crystal modules. Figure \ref{fig:Effect} shows the effect of increasing the density response functions (similarly shaped to those of Fig. \ref{fig:spectral_response}) on the projected light spectrum white LED. Increasing the number of response functions for 30, as in Figure \ref{fig:Effect}, increases the precision of the spectrum projected. The experimental implementation presented here is a system of twenty patterns, but the same experiment can be conducted up to 600 standard sizes, without modification. This may even lead to 1-nm resolution for each strip. Thus, superior performance can be readily achieved.\\
The same procedure was adopted for comparison with the tungsten lamp lighting. The results can be seen in Figures \ref{fig:photonic_standard} and \ref{fig:effect_base-elements}.

\section{Conclusions}

The launch of low-cost and compact spectrophotometers expanded the boundaries of spectrometry well beyond the scientific measures enshrined, for example, transmittance and color comparison. Comparative results were presented for a new compact spectrophotometer, which are based on the properties of photonic crystals arranged in the Trinitron pattern with light coupled through its glass substrate.\\
The response to a light source of a photonic crystal panel pattern on a glass slide is captured by a camera. The resulting image is used to calculate the spectrum of the light source via the response functions of the photonic crystals.\\

The major cost component in the spectrophotometer is the camera. Compact cameras mounted in cell phones and laptops have a very low cost, modular CMOS cameras can be found at retail for less than U\$ 20 \cite{Sparkfun}. The photonic crystal patterns presented here were fabricated using electron beam lithography, nanoimprint lithography retaining the features similar to those of the literature, with excellent uniformity over areas of 3 mm $\times$ 3 mm \cite{Ishihara_et_al}.\\

An advantage of the spectrophotometer based on photonic crystals is that it can be adapted to specific applications by incorporating photonic crystals of particular patterns. This allows the use of cameras with lower resolutions in applications where high resolution is not required in broadband. For example, in a low-cost spectrophotometer designed to detect a specific compound can be taken to the field as the use of batteries, due to its portability and compact size. Further work should search for extending its application to infrared or ultraviolet range. The key would be to test the core photonic crystals that have similar or acceptable response to commercial spectrophotometers with diffraction gratings. Photonic crystals show promise in Telecommunication applications, as can be created photonics devices that drive high-speed, high volume of traffic fiber cables (e.g. the Internet).

\end{document}